\documentclass[aps,prx,twocolumn,superscriptaddress]{revtex4-2}
\usepackage{tabularx}
\usepackage{color,soul}
\usepackage{verbatim}
\usepackage{graphicx}
\usepackage{dcolumn}
\usepackage{amsfonts,amsmath,amssymb,bm}
\usepackage{xcolor}
\usepackage[utf8]{inputenc}
\usepackage[colorlinks=true,allcolors=blue]{hyperref}
\usepackage{amsbsy}  
\usepackage{float}
\usepackage[normalem]{ulem}

\begin{document}
\title{Revealing Charge Transfer in Defect-Engineered 4H$_\text{b}$-TaS$_2$}

\date{\today}
\author{Siavash Karbasizadeh}
\email{siavashk@email.sc.edu}
\affiliation{SmartState Center for Experimental Nanoscale Physics, Department of Physics and Astronomy, University of South Carolina, Columbia, SC, 29208, USA}
\affiliation{Center for Nanophase Materials Sciences, Oak Ridge National Laboratory, Oak Ridge, Tennessee 37831, USA}

\author{Wooin Yang}
\affiliation{Center for Nanophase Materials Sciences, Oak Ridge National Laboratory, Oak Ridge, Tennessee 37831, USA}
\affiliation{Department of Physics and Astronomy, University of Tennessee, Knoxville, Tennessee 37996, USA}

\author{Wonhee Ko}
\affiliation{Department of Physics and Astronomy, University of Tennessee, Knoxville, Tennessee 37996, USA}
\affiliation{Center for Advanced Materials and Manufacturing, University of Tennessee, Knoxville, Tennessee 37920, USA}

\author{Haidong Zhou}
\affiliation{Department of Physics and Astronomy, University of Tennessee, Knoxville, Tennessee 37996, USA}

\author{An-Ping Li}
\affiliation{Center for Nanophase Materials Sciences, Oak Ridge National Laboratory, Oak Ridge, Tennessee 37831, USA}

\author{Tom Berlijn}
\email{berlijnt@ornl.gov}
\affiliation{Center for Nanophase Materials Sciences, Oak Ridge National Laboratory, Oak Ridge, Tennessee 37831, USA}

\author{Sai Mu}
\email{mus@mailbox.sc.edu}
\affiliation{SmartState Center for Experimental Nanoscale Physics, Department of Physics and Astronomy, University of South Carolina, Columbia, SC, 29208, USA}

\begin{abstract}
We present a comprehensive first-principles investigation of defects in 4$H_b$-TaS$_2$. In this layered transition metal dichalcogenide, charge transfer between alternating Mott-insulating 1T and metallic 1H layers gives rise to exotic quantum phases such as the Kondo effect and topological superconductivity. Motivated by recent defect manipulation in 4$H_b$-TaS$_2$ via STM, we address their microscopic nature and impact on interlayer charge transfer. To this end, we systematically analyze over 90 defects using large-scale density functional theory (DFT) calculations. Our analysis identifies 1H sulfur vacancies, Ta-on-S antisites, and Ta interstitials as the most likely origins of the commonly observed defects, with Ta-on-S antisites producing the strongest charge-transfer modulation. Our extensive dataset, compiled from STM simulations, defect formation energies, work functions, and charge transfer, establishes a foundational resource for future theoretical and experimental studies on defect engineering in 4$H_b$-TaS$_2$.
\end{abstract}
\maketitle

\section{INTRODUCTION}\label{I}

\begin{figure}
	\includegraphics[width=\columnwidth]{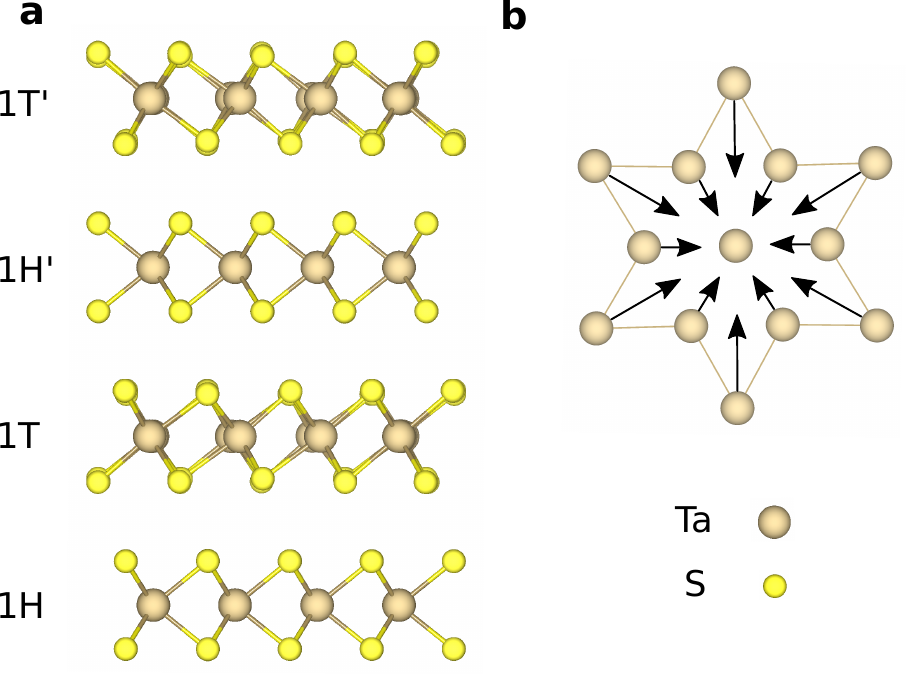}
	\caption{
	(a) Bulk 4$H_b$-TaS$_2$ structure from the side. Brown spheres show tantalum, while the yellow spheres show sulfur. (b) Tantalum atoms constructing the Star-of-David. Black arrows show the displacements of the charge-density-wave distortion. 
	}
	\label{Fig1}
\end{figure}

Transition metal dichalcogenide 4$H_b$-TaS$_2$ is a two-dimensional layered material that consists of alternating layers of 1T-TaS$_2$ and 1H-TaS$_2$ (see Fig. \ref{Fig1}(a)). In bulk 1T-TaS$_2$, the monolayer layers represent a half-filled Mott insulator and a candidate quantum spin liquid\cite{doi:10.1073/pnas.1706769114}. In this layer, clusters of 13 Ta atoms reorganize into a Star-of-David (SoD) configuration, as seen in Fig. \ref{Fig1}(b). The Ta atoms move toward the cluster center, leaving one localized electron on the central Ta atom. This results in localized spins on a triangular lattice, characteristic of the half-filled Mott insulating state \cite{Sipos2008,doi:10.1073/pnas.1706769114,PhysRevB.104.L241114,FAZEKAS1980183,PhysRevB.73.073106,Ruan2021,dong2023,Geng2025,PhysRevB.107.195401}. In contrast, bulk 2H-TaS$_2$, whose layers are equivalent to the 1H layers in 4$H_b$-TaS$_2$, is metallic and becomes a strongly spin-orbit coupled superconductor at Tc = 0.8 K \cite{delaBarrera2018,PhysRevB.98.035203,https://doi.org/10.1002/adma.202305409,PhysRevB.105.L180505,Zhang:gq5013,PhysRevB.27.125}. Combining 1T and 1H polytypes into 4$H_b$-TaS$_2$ offers a rich platform to investigate strongly correlated and topological effects. In this heterostructure, coupling between the localized moments of the 1T layer and the itinerant electrons of the 1H layer significantly modifies the physics of both constituents. The charge transfer depletes the 1T flat band and converts the localized spins into a Kondo lattice, while doping and inversion symmetry breaking in the 1H layers modify the superconducting state into a nodal topological superconductor \cite{doi:10.1073/pnas.2304274120,Shen_2022,PhysRevResearch.6.023224,PhysRevLett.126.256402,wu2025unveilinglandscapemottnessproximity,Date2026,202305017,PhysRevB.102.075138,PhysRevB.109.195170,Ayani2024,Crippa2024,Bae2025,2025-1826,doi:10.1126/sciadv.aax9480}. Scanning tunneling microscopy and transport measurements reveal quasiperiodic one-dimensional modulations and a twofold symmetric superconducting critical field characteristic of a multiple-component nematic superconducting order parameter \cite{Silber2024}. The superconductivity in 4$H_b$-TaS$_2$ is further marked by edge modes and zero-bias conductance peaks at vortex cores consistent with Majorana bound states \cite{Nayak2021}, suggesting relevance for quantum memory and quantum computing. Some more recent results on 4$H_b$-TaS$_2$ also showed that superconductivity competes with an incommensurate lattice-mismatch landscape that modulates charge transfer between the correlated 1T layer and metallic 1H layer, indicating that charge redistribution is intertwined with the superconducting transition \cite{souza2026}. These observations emphasize the crucial role of charge transfer in transforming the Mott-derived localized spins into a Kondo lattice and in enabling the unconventional and topological superconductivity observed in 4$H_b$-TaS$_2$.

\begin{figure}
	\includegraphics[width=\columnwidth]{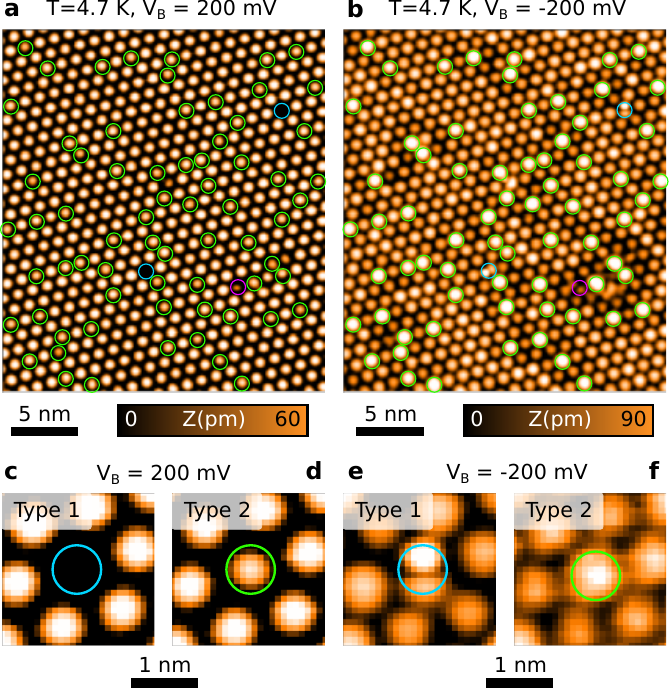}
	\caption{
	Scanning tunneling microscopy (STM) image of 1T-terminated 4$H_b$-TaS$_2$ at $V_\mathrm{bias}$ of (a) 200 mV, and (b) $-$200 mV. Types 1 and 2 defects are specified using cyan and neon green circles. Both images were measured at a tunneling current of 10 pA. Parts (c)-(f) show magnified images of the defects in both positive and negative biases. We have also used a purple circle to show a different type of defect that does not belong in the type 1 or 2 categories.
	}
	\label{Fig2}
\end{figure}

In our recent study \cite{yang2025nanoscalemodulationflatbands}, we demonstrated that charge transfer in 4$H_b$-TaS$_2$ can be further tuned locally through the manipulation of individual defects. In that work, two distinct types of defects were identified, referred to as type 1 and type 2, which can also be visually recognized in earlier scanning tunneling microscopy (STM) data \cite{doi:10.1073/pnas.2304274120}. In the present manuscript, STM images illustrating these two defect types are shown in Figs. \ref{Fig2}(a) and \ref{Fig2}(b), acquired on a 1T-terminated surface. In these images, each bright spot corresponds to a Star-of-David (SoD) cluster, with the bright peak at the cluster center originating from the inward displacement of atoms, as illustrated in Fig. \ref{Fig1}(b). Type 1 defects do not display the central SoD peak at positive bias [see Fig. \ref{Fig2}(c)], whereas type 2 defects retain the SoD peak at positive bias and follow the symmetric SoD pattern, with slightly reduced intensity relative to pristine SoDs, as shown in Fig. \ref{Fig2}(d). At negative bias, type 1 defects exhibit pronounced asymmetry, characterized by an off-centered peak and/or loss of threefold symmetry [Fig. \ref{Fig2}(e)], while type 2 defects show strongly enhanced SoD peak intensity compared to pristine CDW regions [Fig. \ref{Fig2}(f)]. Beyond characterization, STM was also used to reversibly create and remove type ~1 defects, enabling local control of charge transfer and providing a pathway to manipulate the emergent properties of 4$H_b$-TaS$_2$. We should also mention the finding of a different type of defect that does not belong in the type 1 or 2 categories. This defect is highlighted using a purple circle in Fig. \ref{Fig2} and its behavior and origin are currently unknown.

To understand the microscopic origin of the defects shown in Fig. \ref{Fig2}, we previously interpreted them as sulfur vacancies \cite{yang2025nanoscalemodulationflatbands}. Specifically, type ~1 defects were proposed to correspond to sulfur vacancies in the 1T layer, and type ~2 defects to sulfur vacancies in the 1H layer. These assignments are supported by STM simulations and local density of states calculations in that work. While the assignment of type~1 defects to sulfur vacancies is further supported by their ease of STM manipulation, an intriguing observation in the STM data remains unresolved: type~2 defects appear far more frequently than type~1, as seen in previous studies \cite{yang2025nanoscalemodulationflatbands,doi:10.1073/pnas.2304274120} and in our Fig. \ref{Fig2}. The higher prevalence of type~2 defects suggests they may differ in nature, or that a mechanism must be identified to explain their abundance. Furthermore, our STM measurements in Ref.~\cite{yang2025nanoscalemodulationflatbands} revealed that type 1 and type 2 defects qualitatively influence charge transfer between the layers. However, STM primarily resolves lateral spatial variations and can probe only one layer at a time, preventing quantitative determination of the interlayer charge redistribution. These observations motivate a comprehensive DFT study to systematically quantify the charge transfer induced by defects in 4$H_b$-TaS$_2$ and to determine the identity of the abundantly observed type 2 defects. 

In this manuscript, we perform large-scale density functional theory calculations to investigate the impact of defects in 4$H_b$-TaS$_2$ on its interlayer charge transfer, and to study the nature of experimentally observed type ~2 defects. Our study encompasses over 90 defect configurations and includes calculations of defect formation energies, charge transfer, work functions, density of states, and simulated STM images, made publicly available through an online database \cite{data_repository}. By directly comparing STM simulations with experimental observations, we identify three scenarios for the origin of the most abundantly observed type ~2 defects: (i) sulfur vacancies located in the buried 1H layer, (ii) tantalum-on-sulfur antisite defects across the 1T/1H interface, and (iii) Ta interstitials residing between the 1T and 1H layers which exhibit very low formation energies under S-poor conditions. From charge transfer calculations, we find that sulfur vacancies in the 1H layer modify the interlayer charge transfer more strongly than sulfur vacancies in the 1T layer. For Ta$_\mathrm{S}$ antisite defects, we observe particularly strong modulations of charge transfer arising from interlayer bonding between Ta and S atoms across the van der Waals gap, which also stabilizes these defects energetically. From a general perspective, charge transfer between 1T and 1H layers underpins emergent quantum phenomena such as Kondo physics and topological superconductivity, and defects provide a powerful means to locally control this charge transfer. Our extensive simulated dataset thus provides a predictive framework to guide future theoretical and experimental efforts in tailoring the electronic properties of 4$H_b$-TaS$_2$ via defect engineering.

\section{COMPUTATIONAL METHODOLOGY}\label{II}

\begin{figure}
	\includegraphics[width=\columnwidth]{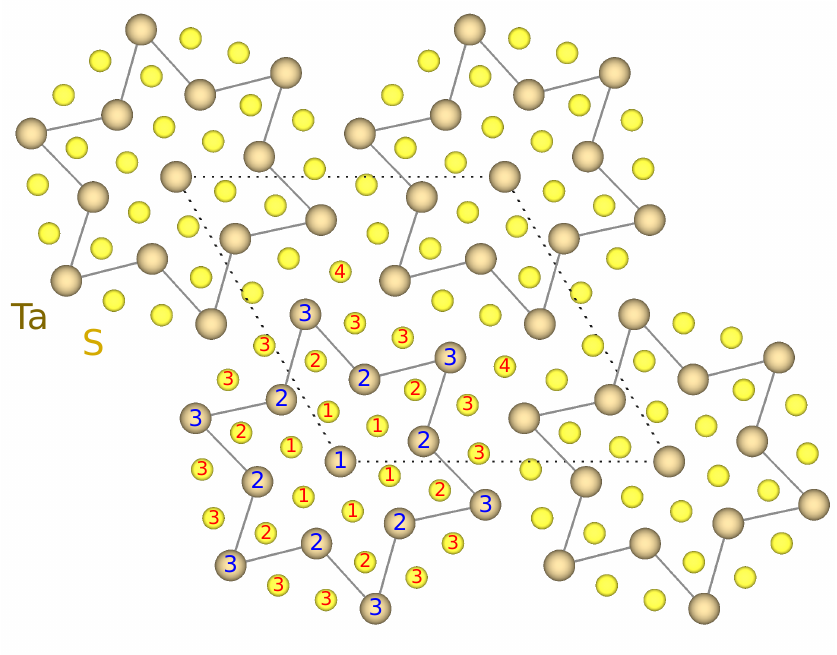}
	\caption{
	1T-TaS$_2$ monolayer from the top. Brown spheres show tantalum, while the yellow spheres show sulfur. The dotted line marks the charge density wave cell considered. The numbers show different lateral positions of sulfur and tantalum atoms with respect to the tantalum in the center of the Star-of-David (Ta1). 
	}
	\label{Fig3}
\end{figure}

The following defects have been considered in our study: sulfur vacancies (V$_\mathrm{S}$), tantalum vacancies (V$_\mathrm{Ta}$), tantalum interstitials between the layers (Ta$_{i1}$), Ta$_\mathrm{S}$ and S$_\mathrm{Ta}$ antisites, as well as oxygen substitutions on sulfur sites (O$_\mathrm{S}$), and iodine substitutions on both sulfur and tantalum sites (I$_\mathrm{S}$ and I$_\mathrm{Ta}$). Most defects were investigated using a 1T/1H bilayer with a $\sqrt{13}\times\sqrt{13}$ in-plane supercell to accommodate the Star-of-David reconstruction in the 1T layer, consisting of an average of 78 atoms. A 12 \AA~vacuum is used for all slab calculations in order to eliminate any spurious interactions with the periodic images. Additionally, we examined the four-layer slabs and $2\sqrt{13}\times2\sqrt{13}$ 1T/1H bilayer supercells for the pristine case, and a few selected defect cases [see Secs. A and B in the Supplemental Material ~\cite{Supp}], indicating similar conclusions to the bilayer 1T/1H system. Figure \ref{Fig3} illustrates all possible lateral positions of the tantalum and sulfur atoms relative to the central tantalum atom of the Star-of-David (Ta1), which are used to define various defects. In total, over 90 supercell calculations were carried out.

First-principles calculations are performed based on density functional theory (DFT) using the Vienna $ab$-$initio$ Simulation Package (VASP) \cite{PhysRevB.48.13115,PhysRevB.54.11169}, employing the projector augmented-wave (PAW) approach \cite{PhysRevB.50.17953,PhysRevB.59.1758}. The exchange-correlation effects are treated under the generalized gradient approximation (GGA) using the Perdew-Burke-Ernzerhof (PBE) functional \cite{PhysRevLett.77.3865}. We also calculated the charge density difference using the spherically averaged GGA+$U$ \cite{PhysRevB.57.1505} ($U_\mathrm{eff}=3$ eV) and GGA+SOC for the pristine 1T/1H bilayer and a few selected defects [see Sec. C in the Supplemental Material ~\cite{Supp}], showing that the difference in calculation details does not alter the qualitative conclusions. The on-site Coulomb interaction of $U_\mathrm{eff}=3$ eV is specifically chosen as the higher limit of the parameter, seeing as previous studies \cite{PhysRevX.7.041054,Lutsyk2023,PhysRevB.90.045134,Lin2020,huang2024ultrafastopticalswitchingheterochiral} have all chosen numbers closer to 2 eV. 

The valence configurations for each element are 3$s^2$3$p^4$ for S and 5$p^6$5$d^4$6$s^1$ for Ta. The nonlocal van der Waals interaction is modeled by the DFT-D3 method \cite{10.1063/1.3382344}. Structural relaxation is performed until the Hellmann-Feynman forces acting on atoms are less than 0.01 eV/\AA. The in-plane lattice parameter was obtained by relaxing the pristine bilayer with the Star-of-David distortion and was found to be 12.06~\AA. This is in good agreement with the experimental observation of $\sqrt{13}\cdot a \approx 12.01$ ~\AA ~\cite{D5TC02933C}. Interlayer distances were relaxed in all simulations and were found to vary only slightly; the full values are summarized in Sec. D2 in the Supplemental Material~\cite{Supp}. The plane-wave energy cutoff is set to 450 eV, and the electronic self-consistency is achieved with a total energy convergence threshold of 10$^{-5}$ eV. A $\Gamma$-centered K-point mesh of $12\times12\times1$ is used for all unitcell calculations, while the $\sqrt{13}\times\sqrt{13}$ SoD CDW structure shown in Fig. \ref{Fig1} is sufficiently converged employing a K-point mesh of $6\times6\times1$. Convergence tests were performed for charge transfer and defect formation energy with respect to K-point mesh and plane-wave cutoff energy [see Sec. D1 in the Supplemental Material ~\cite{Supp}].

All calculations reported in this work were obtained from spin-polarized calculations with the magnetic moments initialized to zero. This choice was made because initializing finite magnetic moments led to spurious magnetization in the 1H layer for the pristine bilayer and even for bulk 2H-TaS$_2$, which is experimentally nonmagnetic. As a practical procedure, we first relaxed each structure with finite initial magnetic moments (5 $\mu_B$) on the Ta atoms and subsequently restarted the calculation from the relaxed structure with all magnetic moments initialized to zero. We found that initializing the moments to 5 $\mu_B$ or zero leads to nearly identical defect formation energies and interlayer charge transfers for representative defects, despite differences in the final magnetic moments. For comparison, Tables S4 and S5 of the Supplemental Material~\cite{Supp} compare calculations initialized with zero and finite magnetic moments using both GGA and GGA+$U$, for selected defects and bulk 2H-TaS$_2$, respectively. Table S4 additionally reports the corresponding defect formation energies and interlayer charge transfers.

To quantify charge transfer (CT) we followed the approach mapped out in Refs. \cite{PhysRevB.110.195138,Wang2018,PhysRevB.109.115112}. Charge density difference ($e$/\AA) can be defined as: 

\begin{equation}
\rho_\mathrm{diff}(z)=\rho_\mathrm{all}(z)-\sum_{i=1}^N\rho_i(z),
\end{equation}

where $\hat{z}$ is chosen as the stacking direction, $\rho_\mathrm{all}(z)$ is the planar-averaged charge density of the full structure, and $\rho_\mathrm{i}(z)$ is the planar-averaged charge density of each constituent; with $i=1,2,\ldots,N$, where $N$ is the number of the aforementioned components. For each constituent, the charge density is computed while maintaining the atomic positions of the full structure. We then integrate $\rho_\mathrm{all}(z)$ in a specified section ($z_1$,$z_2$) to determine the total charge difference:

\begin{equation}
q_\mathrm{diff}(z_1,z_2)=\int_{z_1}^{z_2}\rho_\mathrm{diff}(z)\mathrm{d}z.
\end{equation}

For a given layer, $z_1$ and $z_2$ are chosen as the planes midway between the neighboring layers. Here, the $z$ position of each layer is defined by the average $z$ coordinate of the Ta atoms in that layer. In a bilayer, $z_1$ and $z_2$ for the 1T layer coincide with $z_2$ and $z_1$ for the 1H layer. As a result, the charge gained by 1H is equal in magnitude and opposite in sign to that lost by 1T, so the interlayer charge transfer in the bilayers can be represented by a single value.

For pristine bilayers and for all defect configurations considered above, the bilayer is naturally partitioned into 1T and 1H regions, as the defects can be unambiguously assigned to one of the two layers. An exception arises for Ta interstitial defects, which reside in the interlayer region and cannot be uniquely associated with either layer. In this case, we partition the system into three components: the 1T layer, the 1H layer, and the Ta interstitial. The charge density difference is then defined as $\Delta\rho(z)=\rho_{\mathrm{BL+Ta_i}}(z)-\rho_{\mathrm{1T}}(z)-\rho_{\mathrm{1H}}(z)-\rho_{\mathrm{Ta_i}}(z)$ where $\rho_{\mathrm{Ta_i}}(z)$ is the plane-averaged charge density of an isolated Ta atom computed at the same atomic position. The charge transfer associated with each layer is subsequently obtained by integrating $\Delta\rho(z)$ over the corresponding spatial region, following the same procedure as described above. 

We compared the planar-averaged charge integration approach with Bader charge analysis for select cases (Table S3). Overall, we observed close agreement between the two methods. The primary exception was the Ta$_\mathrm{S1}^\mathrm{1H(top)}$ antisite defect, where Bader analysis yielded a charge transfer of $-$0.38 $e$ compared to $-$0.15 $e$ from our planar-averaged integration approach. This disparity likely arises from out-of-plane relaxation of the Ta$_\mathrm{S1}^\mathrm{1H(top)}$ defect (Fig. \ref{Fig9}(a)). This displacement shifts the zero-flux surface toward the 1T layer, increasing the charge assigned to the 1H layer relative to our approach, which enforces a fixed plane midway between the 1H and 1T layers. In general, charge partitioning schemes are non-unique choices and one method is not inherently superior to another. Here, we favor the planar-averaged charge integration approach because defining a fixed plane midway between the layers is required to quantify the charge-transfer-induced inversion symmetry breaking, which was reported to be a key ingredient for topological superconductivity \cite{doi:10.1126/sciadv.aax9480}.

We define work function as follows:

\begin{equation}
\mathrm{WF}=e V_\mathrm{vac}-E_F,
\end{equation}
where $e$ is the charge of an electron, $V_\mathrm{vac}$ the value of the local potential in the vacuum region far from the slab, and $E_F$ the Fermi level of the slab. We then proceed to define the work function differences using the following formula:

\begin{equation}
\Delta \mathrm{WF}=\mathrm{WF}_\mathrm{1H}-\mathrm{WF}_\mathrm{1T}.
\end{equation}

A positive $\Delta \mathrm{WF}$ indicates charge transfer from the 1T layer to 1H, while a negative value promotes this transfer in the opposite direction. 

To investigate the likelihood of defect incorporation in the structure, we calculate the formation energy $E^f$. For a given defect $X$, the formation energy is defined as: 

\begin{equation}\label{Eq1}
E^f[X]=E_\mathrm{tot}[X]-E_\mathrm{tot}[\mathrm{TaS_2}]-\sum_{i}n_i\mu_i,
\end{equation}
where $E_\mathrm{tot}$ is the total energy from a supercell calculation, $n_i$ is the number of atoms of atomic species $i$ added to ($n_i>0$) or removed from ($n_i<0$) the supercell, and $\mu_i$ is the chemical potential that defines the energy added or lost when exchanging atomic species $i$ with its respective reservoir of atoms. 

The chemical potential $\mu_i$ of each element reflects its relative availability under specific synthesis or processing environments. We define $\mu_i$ in reference to the energy of the corresponding elemental phase, such that $\mu_i=\mu_i^0+\Delta\mu_i$, where $\mu_i^0$ is the total energy of species $i$ in its elemental phase [e.g., solid Ta], and $\Delta\mu_i$ is the deviation from that reference energy. We can then relate values of $\Delta\mu_i$ to the formation enthalpy of TaS$_2$ through thermodynamic equilibrium:

\begin{equation}
\Delta\mu_\mathrm{Ta}+2\Delta\mu_\mathrm{S}=\Delta H_f(\mathrm{TaS_2})
\end{equation}
where $\Delta H_f(\mathrm{TaS_2})$ is calculated to be $-$3.87 eV. For this we used solid elemental Ta and S as references. Two extreme limits can now be determined to span the growth or processing circumstances: S-rich (Ta-poor) conditions, where $\Delta\mu_S=0$; and S-poor (Ta-rich) conditions where $\Delta\mu_S=\frac{1}{2}\Delta H_f(\mathrm{TaS_2})$. Results will be presented in these limiting conditions, while results for intermediate conditions can be obtained using Eq.~\ref{Eq1}. For the chemical potentials of oxygen and iodine, we consider the secondary phases Ta$_2$O$_5$ and TaI$_5$, with the calculated formation enthalpies of $-$20.01 eV and $-$2.75 eV, respectively. A comparison of calculated formation enthalpies and their experimental counterparts is presented in Table \ref{T0}.

\begin{table}[h]
    \centering
\caption{Calculated and reported enthalpies of formation for materials under investigation in the study.}
    \setlength{\tabcolsep}{17pt}
    \begin{tabular}{ccc}
        \hline\hline
        & \multicolumn{2}{c}{$\Delta H_f$ (eV per formula unit)} \\
        \cline{2-3}
        Compound & Calculated & Experimental \\
        \hline
	    TaS$_2$  & $-$3.87 & $-$3.67 \cite{Kubaschewski1993} \\
        Ta$_2$O$_5$  & $-$20.01 & $-$21.12 \cite{https://doi.org/10.1002/adem.202200162} \\
        TaI$_5$  & $-$2.75 & $-$3.04 \cite{https://doi.org/10.1002/zaac.200700530} \\
        \hline\hline
    \end{tabular}
    \label{T0}
\end{table}

STM simulations were visualized using the \textsc{P4VASP} software package based on partial charge density data from VASP. Note that VASP stores the partial charge density in dimensionless units [$\tilde{\rho} = \rho \cdot V$]; therefore, our chosen isosurface value of $0.01$ in the software corresponds to a physical charge density of $3.538 \times 10^{-6}$~\AA$^{-3}$ for the bilayer supercell ($V = 2826.31$~\AA$^{3}$). For the four-layer system ($V = 4694.24$~\AA$^{3}$), the same software input equates to a physical density of $2.130 \times 10^{-6}$~\AA$^{-3}$.

\section{RESULTS AND DISCUSSION}\label{III}

\subsection{Sulfur Vacancies}

In Ref.~\cite{yang2025nanoscalemodulationflatbands} we proposed that type 1 defects are S vacancies in the top 1T layer, and type 2 defects are S vacancies in the buried 1H layer. Good comparisons between STM and DFT were presented in Ref.~\cite{yang2025nanoscalemodulationflatbands} following this scenario, but it leaves open two questions. Why do the STM images in Ref.~\cite{yang2025nanoscalemodulationflatbands,doi:10.1073/pnas.2304274120} and Fig.~\ref{Fig2} show many more type 2 defects  than type 1? And what is their influence on the interlayer charge transfer? We first address these two questions within the S vacancy scenario. 

\subsubsection{Defect Identification via STM-DFT Comparison}

\begin{figure}[!htbp]
	\includegraphics[width=\columnwidth]{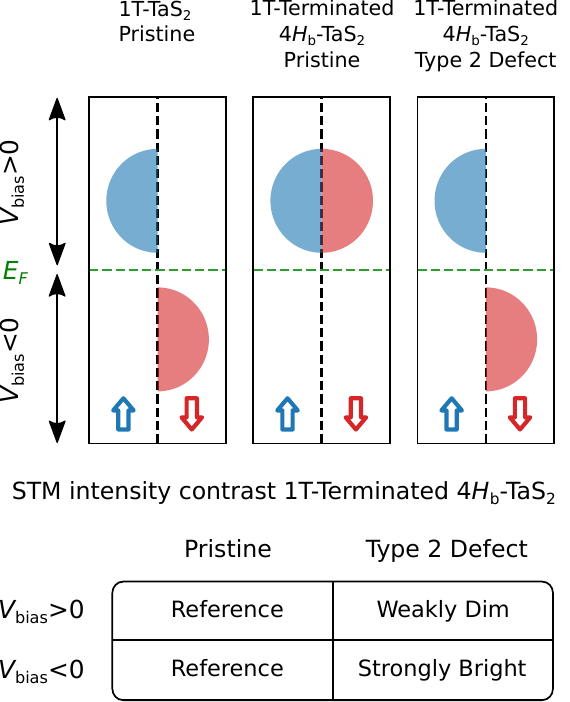}
	\caption{
		Simple picture explaining 1T-terminated 4$H_b$-TaS$_2$ STM contrast of type 2 defects and pristine SoD sites at positive and negative biases, in terms of electron doping Hubbard bands. 1T-TaS$_2$ is a Mott insulator, in pristine 4$H_b$-TaS$_2$ electrons are doped from 1T to 1H, and in the presence of type 2 defect, an electron is doped in 1T, restoring the Mott insulating state. The green dashed line shows the Fermi level. 
	}
	\label{Fig4}
\end{figure}

\begin{figure*}[!htbp]
	\includegraphics*[width=\textwidth]{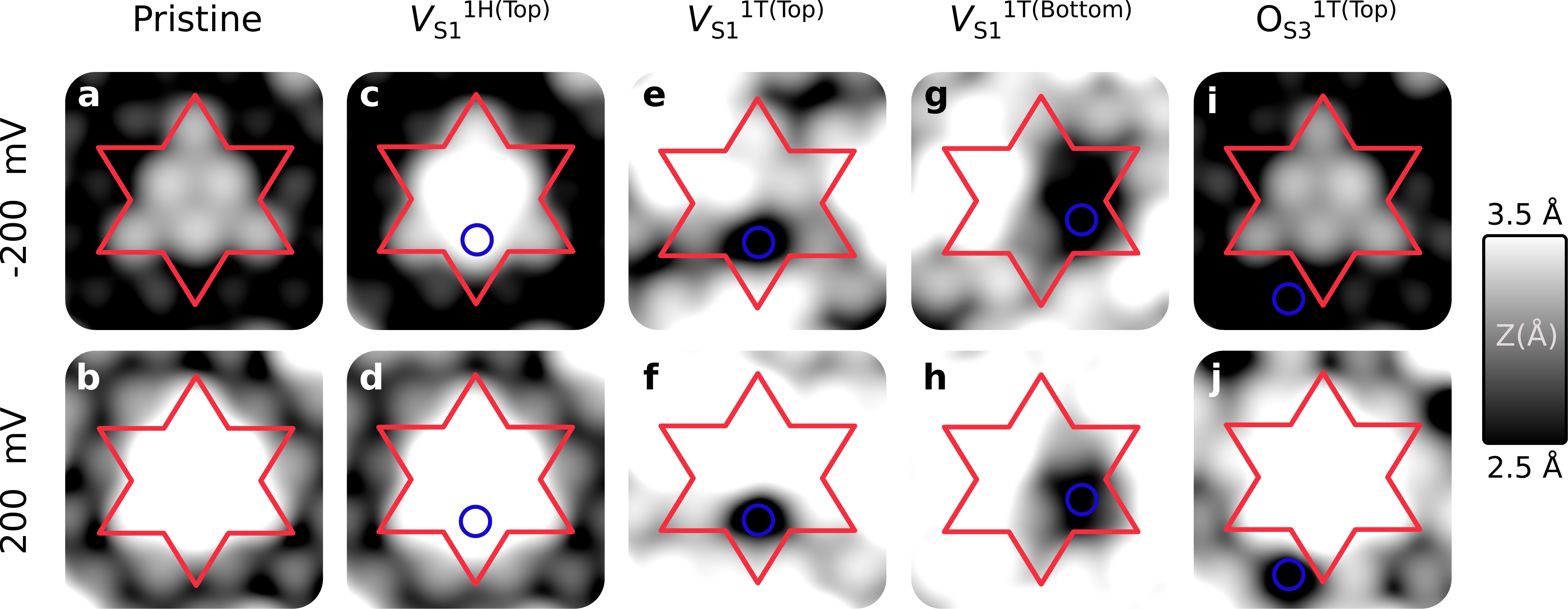}
	\caption{
Scanning tunneling microscopy simulations of the 1T/1H bilayer (a) without defects and with (c) $V_\mathrm{S1}^\mathrm{1H(top)}$, (e) $V_\mathrm{S1}^\mathrm{1T(top)}$, (g) $V_\mathrm{S1}^\mathrm{1T(bottom)}$, and (i) O$_\mathrm{S3}^\mathrm{1T(top)}$ at a bias voltage of $-200$ mV. Panels (b), (d), (f), (h), and (j) show the corresponding simulations for the same configurations at $+200$ mV. The red star and blue circle mark the Star-of-David cluster and the lateral position of the defect, respectively.
	}
	\label{Fig5}
\end{figure*}

Given that type~2 defects are the dominant defect in terms of occurrence, it is useful to begin with a simple picture explaining their STM contrast in terms of electron doping, as illustrated in Fig.~\ref{Fig4}. In 1T-TaS$_2$, the unpaired electrons at the center of each Star-of-David (SoD) form an effectively half-filled Hubbard model, as shown schematically in the left column of the upper panel of Fig.~\ref{Fig4}. In pristine 4$H_b$-TaS$_2$, electrons are transferred from the 1T layer to the 1H layer, resulting in the depopulation of the Hubbard bands in the 1T layer, as shown in the middle column of the upper panel of Fig.~\ref{Fig4}. In contrast, for SoDs hosting type~2 defects, an additional electron is doped back into the 1T layer, filling one of the Hubbard bands and locally restoring a situation similar to that of pristine 1T-TaS$_2$, as illustrated in the right column of the upper panel of Fig.~\ref{Fig4}.

Building on this picture, the STM contrast between pristine SoDs and type~2 defects can be understood using the standard expression for the STM intensity,

\begin{equation}
I(x, y, z) = \int_{E_F}^{E_F+ V_{\text{bias}}} \mathrm{LDOS}(E, x, y, z) dE ,
\end{equation}

where $x,y,z$ are the spatial coordinates, $E_F$ is the Fermi energy, $V_\mathrm{bias}$ is the bias voltage, and LDOS denotes the local density of states. For positive bias, the STM intensity is proportional to the integral of unoccupied states. In pristine SoDs, two empty Hubbard bands contribute to this integral, whereas for type~2 defects only a single empty Hubbard band is present, explaining why type~2 defects appear dim at positive bias. At negative bias, however, pristine SoDs exhibit no occupied Hubbard bands, while type~2 defects host a Hubbard band below the Fermi level, resulting in an enhancement of the STM intensity.

In addition to the opposite contrast at positive and negative biases, the simple picture in Fig.~\ref{Fig2} also explains the experimentally observed asymmetry in magnitude. Type~2 defects appear weakly dim at positive bias but strongly bright at negative bias. This follows naturally from the relative ratios of available states: the ratio of occupied states for type~2 defects to pristine SoDs at negative bias is effectively $1/0$, which is much larger than the corresponding ratio of unoccupied states at positive bias of $2/1$. In addition to reproducing the STM contrast, this simple picture of a half-filled Hubbard band at type~2 defects is further corroborated by scanning tunneling spectroscopy measurements, which reveal Hubbard bands both below and above the Fermi level, as reported in our previous work~\cite{yang2025nanoscalemodulationflatbands} and even more clearly in Ref.~\cite{doi:10.1073/pnas.2304274120}.

Having established a simple picture of type~2 defects as local electron dopants in the 1T layer, we now examine their chemical nature in more detail. Sulfur vacancies are common point defects in transition metal dichalcogenides and act as electron dopant, making them a natural candidate within this framework. In Fig.~\ref{Fig5}, we present first-principles STM simulations for various sulfur-vacancy configurations. To make scanning over many defects manageable, we focus on a single Star-of-David cluster. As a result, comparison between pristine and defect-induced contrast requires separate STM simulations, which are carefully normalized to ensure consistent grayscale height maps. Specifically, identical grayscale values correspond to the same height relative to the surface, defined by the average $z$ position of the top sulfur atoms in the 1T layer. With this procedure, we find that at negative bias the $V_\mathrm{S1}^\mathrm{1H(top)}$ defect reproduces the experimentally observed contrast relative to pristine SoDs: a strong enhancement of brightness appears in contrast to the pristine case. The same qualitative contrast is obtained for sulfur vacancies at the other lateral positions in both the top and bottom sulfur sites of the 1H layer.

So why not assign $V_\mathrm{S}^\mathrm{1T}$ to type~2 defects? These are electron dopants just like $V_\mathrm{S}^\mathrm{1H}$. However, in addition to contrast, another key characteristic of type~2 defects is that the SoD remains mostly centered and threefold symmetric [see Fig.~\ref{Fig2}(d) and (f)]. When a sulfur vacancy resides in the exposed 1T layer, it lies close to the local electronic structure directly probed by STM, naturally leading to strong off-centering and/or threefold-symmetry breaking in the STM map, as shown in Fig.~\ref{Fig5}(e)-(h), regardless of whether $V_\mathrm{S}^\mathrm{1T}$ occupies the top or bottom site of the 1T layer. Such asymmetries are more characteristic of type~1 defects, as shown in Ref.~\cite{yang2025nanoscalemodulationflatbands} and Fig.~\ref{Fig2}(e). This is why in Ref.~\cite{yang2025nanoscalemodulationflatbands} we assigned type~2 defects to $V_\mathrm{S}^\mathrm{1H}$ and type~1 defects to $V_\mathrm{S}^\mathrm{1T}$. Another supporting observation is that in Ref.~\cite{yang2025nanoscalemodulationflatbands} we found that type~1 defects are much easier to manipulate with the STM tip than type~2 defects, consistent with type~1 residing in the exposed 1T layer and type~2 in the buried 1H layer.

\begin{table}[h]
    \centering
\caption{    Formation energies ($E^f$) under sulfur-rich and sulfur-poor conditions, charge transfer (CT), and work function difference ($\Delta$WF) for the 1T/1H bilayer in the presence of sulfur vacancies at different positions relative to the tantalum atom at the center of the Star-of-David [see Fig.~\ref{Fig3}].}
    \setlength{\tabcolsep}{7pt}
    \begin{tabular}{cccccc}
        \hline\hline
        & \multicolumn{2}{c}{$E^f$ (eV)} & & \\
        \cline{2-3}
        Defect & S-rich & S-poor & CT ($e$) & $\Delta$WF (eV) \\
        \hline
		Pristine & N/A & N/A & 0.35 & 0.81 \\
        $V_\mathrm{S1}^\mathrm{1T(top)}$    & 2.47 & 0.54 & 0.38 & 0.89 \\
        $V_\mathrm{S2}^\mathrm{1T(top)}$    & 2.47 & 0.54 & 0.38 & 0.89 \\
        $V_\mathrm{S3}^\mathrm{1T(top)}$    & 2.43 & 0.50 & 0.38 & 0.89 \\
        $V_\mathrm{S4}^\mathrm{1T(top)}$    & 2.48 & 0.55 & 0.37 & 0.89 \\ 
        $V_\mathrm{S1}^\mathrm{1T(bottom)}$ & 2.70 & 0.77 & 0.34 & 0.90 \\ 
        $V_\mathrm{S2}^\mathrm{1T(bottom)}$ & 2.59 & 0.66 & 0.35 & 0.91 \\ 
        $V_\mathrm{S3}^\mathrm{1T(bottom)}$ & 2.63 & 0.70 & 0.35 & 0.93 \\ 
        $V_\mathrm{S4}^\mathrm{1T(bottom)}$ & 2.65 & 0.72 & 0.35 & 0.91 \\ 
        $V_\mathrm{S1}^\mathrm{1H(top)}$    & 2.46 & 0.53 & 0.23 & 0.57 \\ 
        $V_\mathrm{S2}^\mathrm{1H(top)}$    & 2.51 & 0.58 & 0.24 & 0.59 \\ 
        $V_\mathrm{S3}^\mathrm{1H(top)}$    & 2.49 & 0.56 & 0.24 & 0.59 \\ 
        $V_\mathrm{S4}^\mathrm{1H(top)}$    & 2.46 & 0.53 & 0.24 & 0.58 \\ 
        $V_\mathrm{S1}^\mathrm{1H(bottom)}$ & 2.39 & 0.46 & 0.23 & 0.58 \\ 
        $V_\mathrm{S2}^\mathrm{1H(bottom)}$ & 2.44 & 0.51 & 0.25 & 0.59 \\ 
        $V_\mathrm{S3}^\mathrm{1H(bottom)}$ & 2.41 & 0.48 & 0.25 & 0.60 \\ 
        $V_\mathrm{S4}^\mathrm{1H(bottom)}$ & 2.40 & 0.47 & 0.24 & 0.60 \\ 
        \hline\hline
    \end{tabular}
    \label{T1}
\end{table}

Despite the consistencies between the experimental observations and the sulfur vacancy scenario, there remain two issues. The first is the question we already raised: why are there many more type~2 defects compared to type~1 defects? In Table~I, we present the formation energies of $V_\mathrm{S}$ under S-rich and S-poor conditions. The data reveals that vacancies in the 1T(bottom) layer possess formation energies roughly 200~meV higher than those in the 1H or 1T(top) layers, providing a thermodynamical basis for their scarcity. This is consistent with the observation that type~1 defects are easily manipulated by the STM tip, suggesting they are located on the exposed top surface rather than buried. However, 1T(top) energies are comparable to—and sometimes lower than—those of the 1H layer. Furthermore, from a kinetics point of view, sulfur can escape into vacuum from the 1T layer, whereas it remains trapped in 1H beneath the 1T layer. To explain the lack of observed 1T(top) defects, we turn to Ref.~\cite{Barja2019}, which notes that oxygen-chalcogen substitutions can be prolific in TMDs. Even in ultra-high vacuum, the top 1T layer is inevitably exposed to dilute O gas. Since S and O are isovalent, such substitutions are energetically favorable and can ``heal'' the local electronic structure. Comparing Fig.~\ref{Fig5}(i)(j) against (a)(b) shows that these (S,O) substitutions strongly mimic pristine SoDs. This could explain why fewer sulfur vacancies are observed in the 1T(top) layer compared to 1H: O substitutions effectively obscure the STM signature of 1T vacancies. We note that for some lateral positions these substitutions shift the peak away from the SoD center while preserving threefold symmetry [see Fig.~S15 in the Supplemental Material~\cite{Supp}]. We attribute this to the artifact of simulating periodic rather than dilute, random substitutions. For dilute distributions, we expect the surrounding pristine SoDs to restore the peak in the defected SoD to its original center.

The second issue with interpreting type~2 defects as sulfur vacancies in 1H is that, while the strong brightness contrast at negative bias is reproduced, the corresponding dimness at positive bias is not captured in the simulations shown in Fig.~\ref{Fig5}(a)-(d). We note that this contrast is intrinsically more difficult to reproduce, since it is a weaker effect than the pronounced brightness at negative bias, consistent with the discussion above in terms of the relative ratios of available states at positive and negative bias. 

\subsubsection{Interlayer Charge Transfer Modulation}

One of the exciting possibilities of defect manipulation reported in Ref.~\cite{yang2025nanoscalemodulationflatbands} is that it opens the door to local tuning of charge transfer, which in turn can be used to modify the exotic properties of 4$H_b$-TaS$_2$. While such charge transfer effects are qualitatively suggested by STM in Ref.~\cite{yang2025nanoscalemodulationflatbands}, quantitatively extracting these via this technique is difficult. Although changes in local charge influence the STM height, the measured signal also incorporates effects from electronic wave functions and atomic structure, complicating quantitative interpretation. Therefore, to directly quantify how S vacancies modify the interlayer charge transfer, we perform DFT calculations.

\begin{figure}
	\includegraphics[width=\columnwidth]{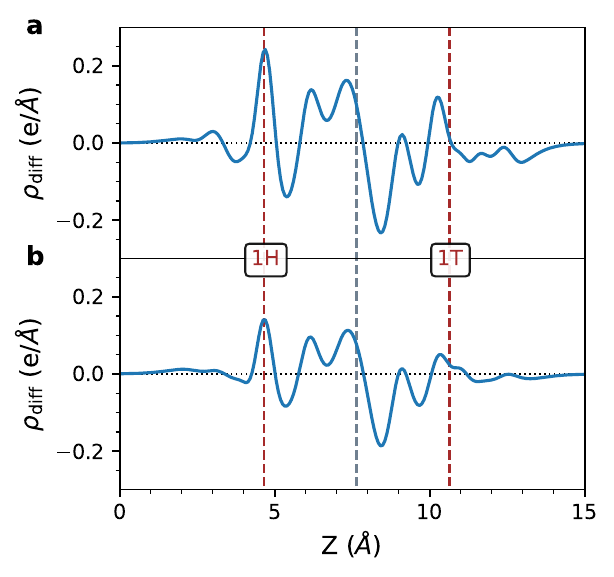}
	\caption{
		Charge density difference for (a) pristine 1T/1H bilayer and (b) 1T/1H bilayer with $V_\mathrm{S1}^\mathrm{1H(top)}$, a sulfur vacancy in the top of the 1H layer. Brown dashed lines show the average height of Ta atoms for each layer, while the grey dashed line is the midpoint between the two layers. 
	}
	\label{Fig6}
\end{figure}

\begin{figure*}[!htbp]
	\includegraphics*[width=\textwidth]{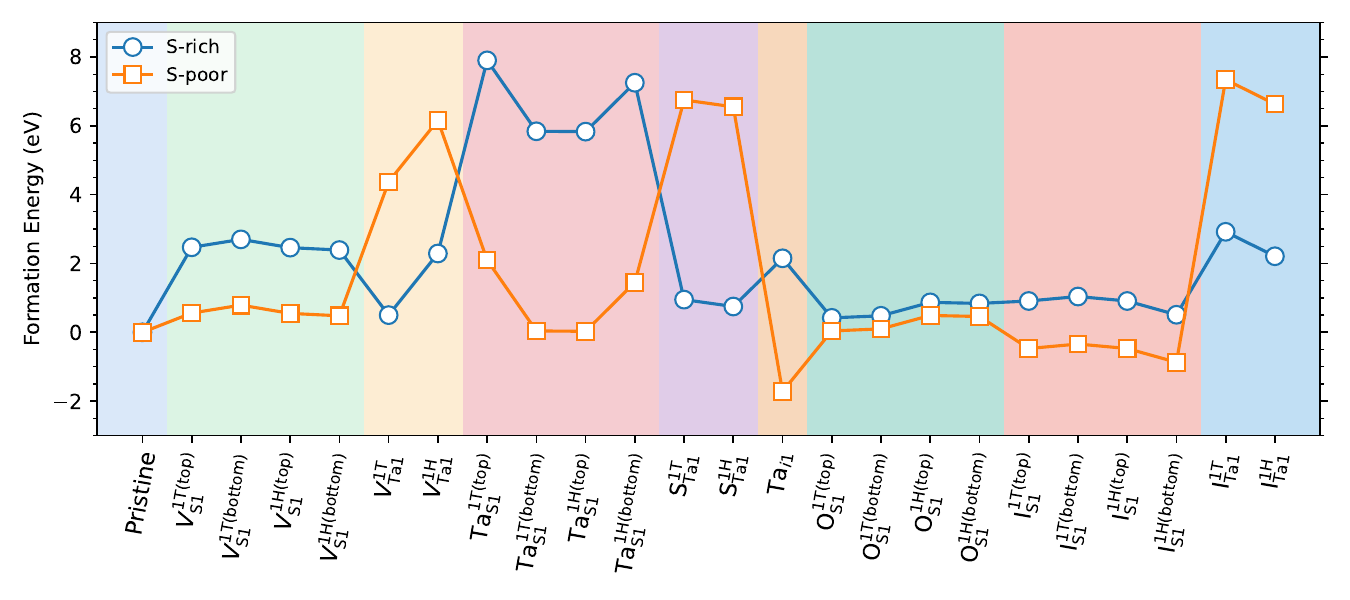}
	\caption{
Formation energies in S-poor and S-rich conditions for pristine 1T/1H bilayers and bilayers with various single defects. We show one representative lateral position per vertical site, as vertical position dominates the formation energy.	}
	\label{Fig7}
\end{figure*}

To quantify the charge transfer modulation, we compute the difference between the charge of the full bilayer and the charges of the individual layers, following the procedure of Ref.\cite{PhysRevB.110.195138,Wang2018,PhysRevB.109.115112}. We first planar-average this difference and then integrate it within the 1T and 1H regions. These regions are defined by planes midway between the 1T and 1H layers, which in turn are determined from the average Ta $z$ positions [see Sec. \ref{II} for more details]. In Fig.~\ref{Fig6} we show the results for the pristine bilayer and for a sulfur vacancy in the top of the 1H layer, $V_\mathrm{S1}^\mathrm{1H(top)}$. In Fig.~\ref{Fig6}(a), the planar-averaged density difference in units of electron charge is mostly positive in the 1H region and mostly negative in the 1T region, representing the charge transfer from 1T to 1H. In Fig.~\ref{Fig6}(b), these positive and negative fluctuations are reduced, demonstrating that the sulfur vacancy in 1H suppresses the charge transfer from 1T to 1H. This is expected, since sulfur vacancies in 1H effectively donate extra electrons that block the charge transfer from 1T. This finding is also consistent with the simple picture described in Fig.~\ref{Fig4}, in which sulfur vacancies in 1H act like type~2 defects that inhibit charge transfer and restore the central 1T Ta-$d{z^2}$ electrons to their half-filled Hubbard band state.

\begin{figure*}[!htbp]
	\includegraphics*[width=\textwidth]{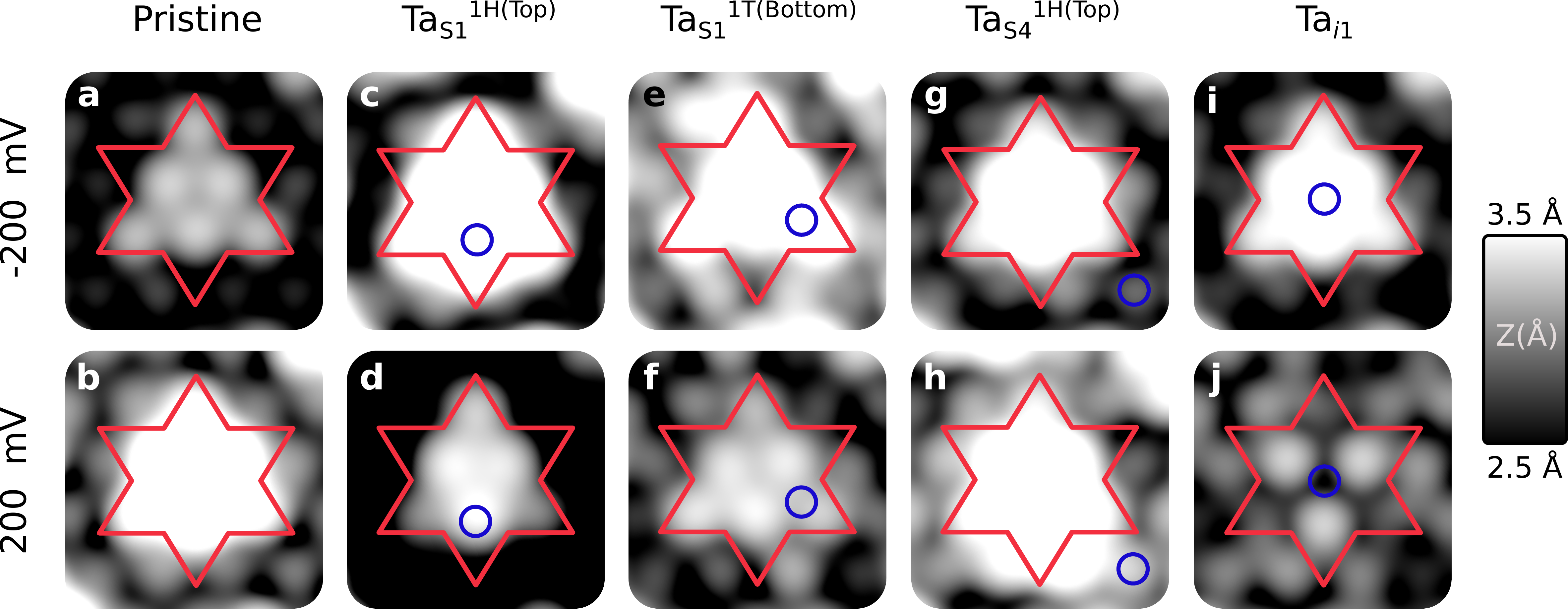}
	\caption{
Scanning tunneling microscopy simulations of the 1T/1H bilayer (a) without defects and with Ta on S antisite defects (c) Ta$_\mathrm{S1}^\mathrm{1H(top)}$, (e) Ta$_\mathrm{S1}^\mathrm{1T(bottom)}$, (g) Ta$_\mathrm{S4}^\mathrm{1H(top)}$, and Ta interstitial (i) Ta$_{i1}$ at a bias voltage of $-200$ mV. Panels (b), (d), (f), (h), and (j) show the corresponding simulations for the same configurations at $+200$ mV. The red star and blue circle mark the Star-of-David cluster and the lateral position of the defect, respectively.
	}
	\label{Fig8}
\end{figure*}

If we integrate the planar-averaged charge density modulations in Fig.~\ref{Fig6} over the 1H region, we obtain values 0.35 $e$ and 0.23 $e$ for the pristine bilayer, and the bilayer with a sulfur vacancy in 1H, respectively. By definition, the opposite values are obtained when integrating over the 1T region. This shows that the sulfur vacancy in 1H reduces the interlayer charge transfer by 0.12 $e$. In Table~\ref{T1}, we list the numerical values for the charge transfers for all S vacancies. Corresponding planar-averaged density difference plots are shown in Fig. S21 in the Supplemental Material~\cite{Supp}. Two points are noteworthy. First, the lateral position of the vacancy does not significantly influence the charge transfer. On the other hand, there is a substantial difference in charge transfer modulation depending on whether the vacancy resides in 1T or 1H. The sign of the modulation is opposite, as expected: placing a vacancy in 1T effectively injects an extra electron into 1T, which enhances the charge transfer. The less expected observation is that the magnitude of the charge transfer modulation induced by a sulfur vacancy in 1T is roughly an order of magnitude smaller than that by a sulfur vacancy in 1H.

We are not entirely certain why there is a pronounced asymmetry in the magnitude of the charge transfer modulation between vacancies in 1T and 1H. However, we have identified two clues. First, the charge transfer correlates with the difference in work function between the two layers. For the pristine bilayer, the work functions computed for the individual 1T and 1H layers are 5.11~eV and 5.93~eV, respectively, giving a difference of 0.81~eV. This misalignment of Fermi levels is consistent with the direction of charge transfer from 1T to 1H. As shown in Table~\ref{T1}, sulfur vacancies in 1H reduce the work-function difference between 1T and 1H by about 0.22--0.24~eV, whereas sulfur vacancies in 1T increase the work function by only $\sim$0.08--0.12~eV. This is mirroring the trend observed for the charge transfer in terms of both sign and magnitude. Second, if the $V_\mathrm{S1}^{\mathrm{1H}}$ induced atomic relaxations are excluded from our 1T/1H bilayer simulation, the charge transfer becomes 0.31~$e$, significantly closer to the pristine value than in the fully relaxed case (0.23~$e$). This shows that structural relaxation plays an important role in the larger change in charge transfer caused by 1H vacancies compared to 1T vacancies.

\subsection{Other Defects}

In Ref.~\cite{yang2025nanoscalemodulationflatbands} and in the discussion in this manuscript so far, we have proposed that type~2 defects can be explained in terms of sulfur vacancies. However, many other native defects are possible. We therefore also consider Ta vacancies, antisite defects (Ta$_\mathrm{S}$ and S$_\mathrm{Ta}$) at all lateral positions, as well as Ta interstitials located at various lateral positions between the 1T and 1H layers. In addition, we examine iodine substitutions on S and Ta sites, motivated by the use of iodine as a transport agent in the chemical vapor transport growth of 4$H_b$-TaS$_2$. Fig.~\ref{Fig7} provides an overview of the formation energies of these defects, where a representative lateral position is shown for each defect type, as lateral variations are comparatively small. The full results are presented in Tables~\ref{T1} and~\ref{T2} and Sec. D4 of the Supplemental Material~\cite{Supp}. We emphasize that while formation energies offer useful guidance, they do not provide a complete picture of defect populations in chemically vapor–transport–grown 4$H_b$-TaS$_2$, as the growth process is inherently non-equilibrium and influenced by chemical reactions, gas-phase chemical potentials, transport gradients, and kinetic barriers. To further identify which of these defects are realized experimentally, we therefore turn again to a comparison between DFT simulations and STM measurements.

\subsubsection{Defect Identification via STM-DFT Comparison}

Among the other native defects besides sulfur vacancies, Ta-on-S antisites are an interesting possibility because, like sulfur vacancies, they act as electron dopants. Interestingly, we see from Fig.~\ref{Fig8}(a)-(d) that some Ta-on-S antisites in the top of the 1H layer, such as Ta$_\mathrm{S1}^\mathrm{1H(top)}$, produce the correct contrast at both positive and negative bias. However, some other Ta-on-S antisite defects, such as those located in the bottom of 1T [e.g., Ta$_\mathrm{S2}^\mathrm{1T(bottom)}$], do not reproduce the STM image correctly. These defects appear bright compared to pristine at positive bias and, notably, are strongly off-centered relative to the SoD center and/or lack threefold symmetry [see Fig. S11 in the Supplemental Material~\cite{Supp}]. Not all defects in 1T produce a poor match, though. For example, Ta$_\mathrm{S1}^\mathrm{1T(bottom)}$, shown in Fig.~\ref{Fig8}(e)(f), exhibits the correct contrast and a peak located at the center of the SoD. Conversely, some Ta-on-S antisites in 1H top, such as Ta$_\mathrm{S4}^\mathrm{1H(top)}$, display the wrong contrast compared to experiment. Although centered closely to the SoD, it appears bright rather than dim relative to pristine SoD at positive bias, as shown in Fig.~\ref{Fig8}(b)-(h).

\begin{table}
    \caption{     
        Formation energies ($E^f$) under sulfur-rich and sulfur-poor conditions, charge transfer (CT), and work function difference ($\Delta$WF) for the 1T/1H bilayer in the presence of Ta on S antisite defects and Ta interstitials at different positions relative to the tantalum atom at the center of the Star-of-David [see Fig.~\ref{Fig3}].
		The lowest formation energies are highlighted in bold; these defects reproduce the characteristic STM contrasts compared to pristine Stars-of-David, as observed in STM experiments [compare Fig.~\ref{Fig8}(a)-(f) and (i)(j) against Fig.~\ref{Fig2}(d)(f)].
    }
    \setlength{\tabcolsep}{7pt}
    \begin{tabular}{cccccc}
        \hline\hline
        & \multicolumn{2}{c}{$E^f$ (eV)} & & \\
        \cline{2-3}
        Defect & S-rich & S-poor & CT ($e$) & $\Delta$WF (eV) \\
        \hline
        Ta$_\mathrm{S1}^\mathrm{1T(top)}$    & 7.88 & 2.08 & 0.27 & 0.67 \\
        Ta$_\mathrm{S2}^\mathrm{1T(top)}$    & 7.85 & 2.05 & 0.29 & 0.74 \\
        Ta$_\mathrm{S3}^\mathrm{1T(top)}$    & 7.90 & 2.10 & 0.28 & 0.71 \\
        Ta$_\mathrm{S4}^\mathrm{1T(top)}$    & 7.85 & 2.05 & 0.29 & 0.70 \\ 
$\mathbf{Ta}_{\mathbf{S1}}^{\mathbf{1T(bottom)}}$ & $\mathbf{5.82}$ & $\mathbf{0.02}$ & $\mathbf{0.65}$ & $\mathbf{0.86}$ \\
		Ta$_\mathrm{S2}^\mathrm{1T(bottom)}$ & 5.94 & 0.14 & 0.60 & 0.85 \\ 
        Ta$_\mathrm{S3}^\mathrm{1T(bottom)}$ & 5.95 & 0.15 & 0.60 & 0.85 \\ 
        Ta$_\mathrm{S4}^\mathrm{1T(bottom)}$ & 5.94 & 0.14 & 0.60 & 0.85 \\
$\mathbf{Ta}_{\mathbf{S1}}^{\mathbf{1H(top)}}$ & $\mathbf{5.81}$ & $\mathbf{0.01}$ & $\mathbf{-0.15}$ & $\mathbf{0.43}$ \\
        Ta$_\mathrm{S2}^\mathrm{1H(top)}$    & 6.08 & 0.28 & $-$0.18 & 0.23 \\
        Ta$_\mathrm{S3}^\mathrm{1H(top)}$    & 5.87 & 0.07 & $-$0.13 & 0.44 \\
        Ta$_\mathrm{S4}^\mathrm{1H(top)}$    & 6.32 & 0.52 & $-$0.03 & 0.17 \\
        Ta$_\mathrm{S1}^\mathrm{1H(bottom)}$ & 7.23 & 1.43 & 0.24 & 0.65 \\ 
        Ta$_\mathrm{S2}^\mathrm{1H(bottom)}$ & 7.24 & 1.44 & 0.25 & 0.66 \\ 
        Ta$_\mathrm{S3}^\mathrm{1H(bottom)}$ & 7.24 & 1.44 & 0.25 & 0.66 \\ 
        Ta$_\mathrm{S4}^\mathrm{1H(bottom)}$ & 7.23 & 1.43 & 0.25 & 0.66 \\
				        \hline
$\mathbf{Ta}_{\mathbf{i1}}$ & $\mathbf{2.13}$ & $\mathbf{-1.74}$ & $\mathbf{0.33}$ & $\mathbf{0.72}$ \\
Ta$_{i2}$                  & 2.32 & $-$1.55 & 0.31 & 0.70 \\
Ta$_{i3}$                  & 2.25 & $-$1.62 & 0.26 & 0.63 \\
        \hline\hline
    \end{tabular}
    \label{T2}
\end{table}

\begin{figure}
	\includegraphics[width=\columnwidth]{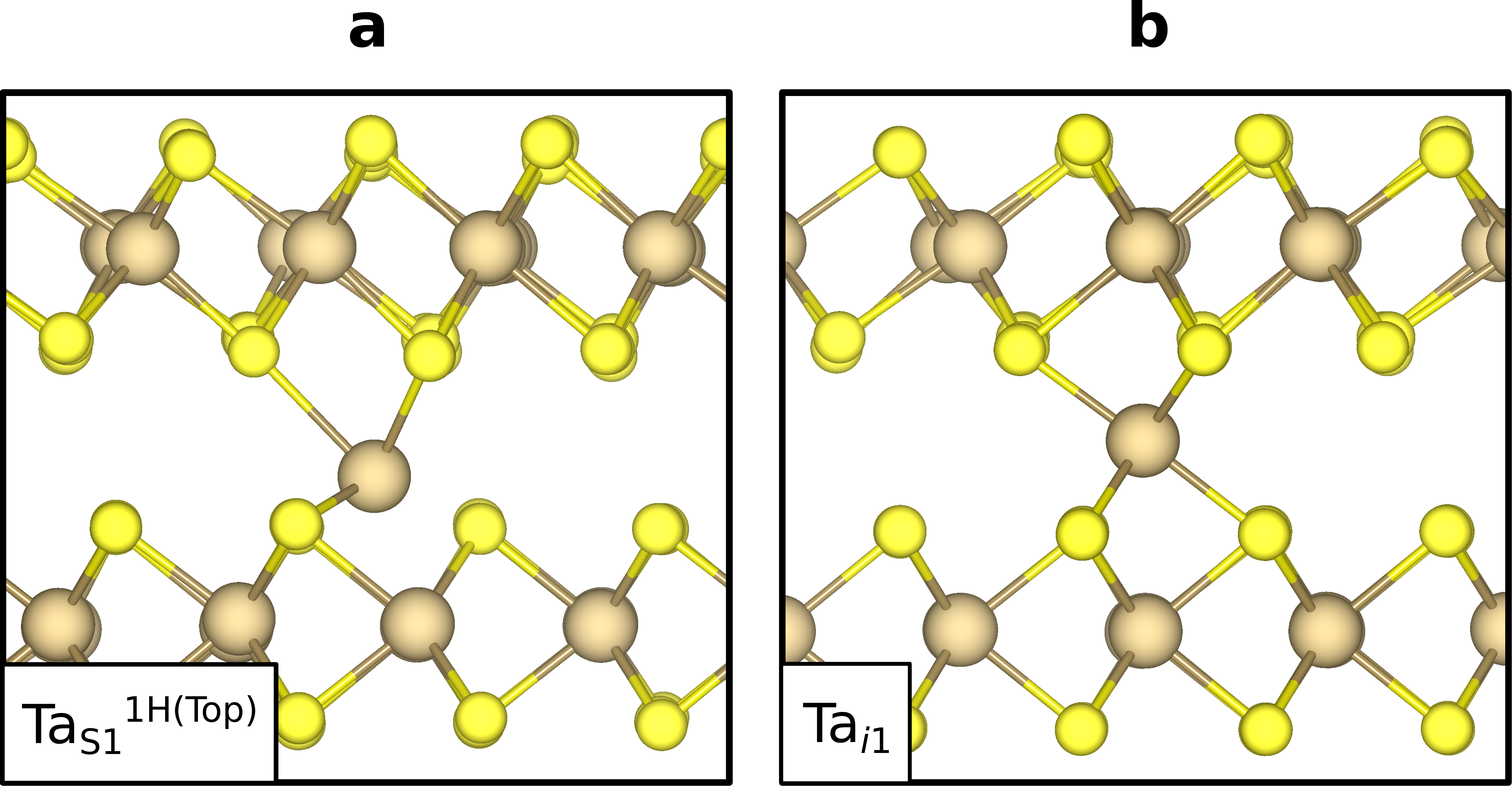}
	\caption{
		Structures of 1T/1H bilayer with (a) Ta on S antisite Ta$_\mathrm{S1}^\mathrm{1H(top)}$ displaying bonding with the bottom 1T S atoms, and (b) the Ta interstitial Ta$_{i1}$. Brown spheres show tantalum, while the yellow spheres show sulfur. 
	}
	\label{Fig9}
\end{figure}

To better understand how the experimentally observed STM contrast compares with the Ta$_\mathrm{S}$ antisite simulations, we next turn to their defect formation energies listed in Table~\ref{T2}, which include the full set of lateral configurations for each vertical site. While formation energies alone cannot determine defect populations in chemically vapor–transport–grown 4$H_b$-TaS$_2$, owing to non-equilibrium growth conditions and kinetic effects, they provide a more stringent criterion when comparing defects of the same type. We therefore proceed to use the formation energies to assess the relative likelihood of different Ta$_\mathrm{S}$ configurations. We observe a clear trend: Ta$_\mathrm{S}$ antisites located between the layers [bottom 1T or top 1H] have significantly lower formation energies than Ta$_\mathrm{S}$ antisites at the exposed surfaces of the bilayer [top 1T and bottom 1H]. We find that this energetic preference arises because a Ta$_\mathrm{S}$ at the bottom of the 1T layer can move a bit out of the layer and bond with sulfur atoms in the top of the 1H layer, and vice versa. This bonding configuration is illustrated in the side-view structural schematic shown in Fig.~\ref{Fig9}(a).

We thus find that Ta$_\mathrm{S}$ antisites located between the 1T and 1H layers have significantly lower formation energies, due to the energy gained from Ta bonding with sulfur atoms in the opposing 1T or 1H layer. Nonetheless, this alone does not explain the observed STM intensities, because not all Ta$_\mathrm{S}$ defects between the layers reproduce the correct STM contrast. For example, in Fig.~\ref{Fig8}(g)(h) we see that Ta$_\mathrm{S4}^\mathrm{1H(top)}$ displays the wrong contrast at positive bias. However, among the Ta$_\mathrm{S}$ defects located between the layers, those with the lowest formation energies are precisely the ones that yield the correct STM contrast: in the S-poor limit, Ta$_\mathrm{S1}^\mathrm{1H(top)}$ and Ta$_\mathrm{S1}^\mathrm{1T(bottom)}$ have formation energies of 0.01~eV and 0.02~eV, respectively. These defects are highlighted in bold in Table~\ref{T2}. The next higher-energy defect, Ta$_\mathrm{S3}^\mathrm{1H(top)}$ with a formation energy of 0.07~eV in the S-poor limit, also reproduces the correct STM contrast, albeit with a slightly off-centered peak relative to the Star-of-David [see Fig. S12 in the Supplemental Material~\cite{Supp}]. In contrast, Ta$_\mathrm{S4}^\mathrm{1T(bottom)}$ has a formation energy of 0.14~eV in the S-poor limit and exhibits the incorrect contrast at positive bias [see Fig. S11 in the Supplemental Material~\cite{Supp}]. Thus, in the S-poor limit, Ta$_\mathrm{S}$ defects that produce the wrong STM contrast have formation energies at least seven times higher than those that reproduce the experimentally observed contrast, which may explain why such defects are not observed experimentally.

In addition to Ta-on-S antisites, Ta interstitials are also electron dopants and are therefore interesting candidates for the type~2 defects. By comparing Fig.~\ref{Fig8}(a,b) and Fig.~\ref{Fig8}(i,j), we find that the Ta$_{i1}$ interstitial [located above Ta$_1$, as defined in Fig.~\ref{Fig3}] exhibits the correct STM contrast, appearing bright (dim) at negative (positive) bias, with a peak centered on the SoD. In contrast, Ta interstitials at other lateral positions [Ta$_{i2}$ and Ta$_{i3}$] show the wrong contrast and display peaks that are off-centered relative to the SoD and/or not threefold symmetric. As in the case of Ta$_\mathrm{S}$ defects, we find that the Ta interstitial exhibiting the correct STM contrast also has the lowest formation energy [highlighted in bold in Table~\ref{T2}], whereas the higher-energy Ta interstitials correspond to the incorrect STM signatures. This energetic ordering may therefore explain why Ta$_{i2}$ and Ta$_{i3}$ are not observed experimentally, particularly under growth conditions intermediate between S-rich and S-poor, where the formation energy of Ta$_{i1}$ is close to zero. We note that Ta interstitials below S sites were also considered but were found to be unstable or to have prohibitively high formation energies.

\begin{figure*}[!htbp]
	\includegraphics*[scale=0.8]{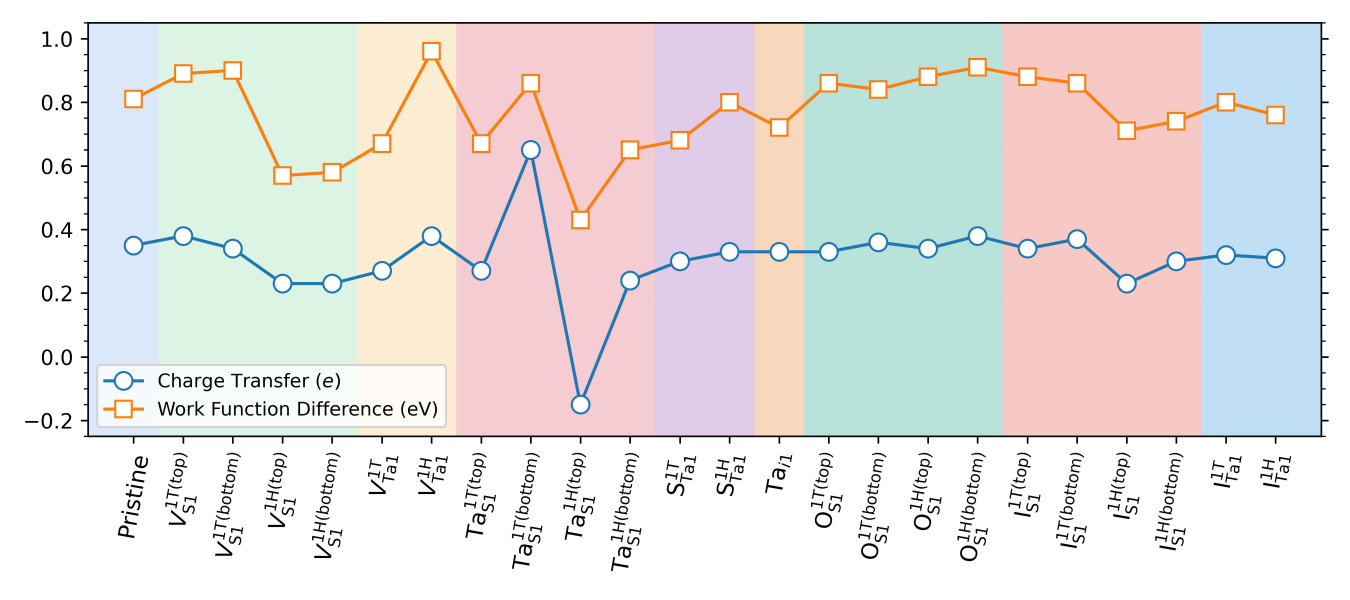}
	\caption{
 Comparison of work function difference and charge transfer for pristine 1T/1H bilayers and bilayers with various single defects. Only one defect per vertical position is shown, as lateral displacement has little effect. Outliers significantly higher than the lateral average are O$_\mathrm{S3}^\mathrm{1T(top)}$ and O$_\mathrm{S4}^\mathrm{1T(top)}$ (see Tables~\ref{T1} and~\ref{T2} and Tables S10-S12 in the Supplemental Material~\cite{Supp}).	}
	\label{Fig10}
\end{figure*}

We also performed STM simulations for other native defects, including S$_\mathrm{Ta}$ antisites and Ta vacancies. However, none of these defects show good agreement with the experimental STM images, either in terms of the bias-dependent contrast or the position of the intensity maximum [see Sec. C2 in the Supplemental Material~\cite{Supp}]. In addition, we considered iodine substitutions on sulfur sites, motivated by the use of iodine as a transport agent in the chemical vapor transport growth of the crystal. Since I for S substitutions act as electron dopants, they could in principle be suitable candidates for type~2 defects. However, their simulated STM signatures do not reproduce the experimental observations [see Fig. S17 in the Supplemental Material~\cite{Supp}]. This leaves Ta$_\mathrm{S}$ antisites and Ta interstitials as the most plausible candidates for type 2 defects, in addition to the previously discussed sulfur vacancies.

\subsubsection{Interlayer Charge Transfer Modulation}

To provide an overview of how different defects can be used to engineer the interlayer charge transfer in 4$H_b$-TaS$_2$, we summarize in Fig.~\ref{Fig10} the charge transfer induced by all defects relative to the pristine bilayer, together with the corresponding work-function differences. Corresponding planar-averaged density difference plots are shown in Sec. D5 in the Supplemental Material~\cite{Supp}, while the correlation between charge transfer and work function difference is quantified in Fig. S29. The most striking feature is that Ta$_\mathrm{S}$ antisites located between the layers induce exceptionally large modulations of the interlayer charge transfer, with a strong dependence on whether the defect resides in the 1T or 1H layer. A Ta$_\mathrm{S}$ antisite in 1T roughly doubles the charge transfer, increasing it from about 0.35 to 0.7 electrons per Star-of-David cluster, whereas a Ta$_\mathrm{S}$ antisite in 1H reverses the direction of charge transfer, with approximately 0.1 electrons flowing from 1H to 1T. The origin of these pronounced charge-transfer modulations can again be traced to the bonding between the Ta$_\mathrm{S}$ antisite and sulfur atoms in the opposite layer, as illustrated in Fig.~\ref{Fig9}(a). Through these bonds, a Ta$_\mathrm{S}$ antisite in 1H injects electrons into the 1T layer, and vice versa for a Ta$_\mathrm{S}$ antisite in 1T. Consequently, the strong charge-transfer effects and the low formation energies of these defects share a common physical origin.

Another remarkable feature of Fig.~\ref{Fig10} is the strong correlation between the charge transfer and the work-function difference. This indicates that, at least to some extent, the interlayer charge transfer is governed by electrostatic alignment arising from the work-function mismatch between the individual 1T and 1H layers, rather than being dominated purely by interlayer hybridization or interactions. Defects locally modify the intrinsic electronic properties of each layer, and the resulting charge transfer reflects these layer-resolved changes. Interlayer coupling still plays an important role, particularly for Ta$_\mathrm{S}$ antisites and Ta interstitials where strong bonding occurs between opposite layers, but the correlation between the charge transfer and the work-function difference in Fig.~\ref{Fig10} shows that individual layer energetics constitute a key driving force.

\subsection{Hybridization Effects on Charge Transfer}

\begin{figure}[!htbp]
	\includegraphics[scale=0.66]{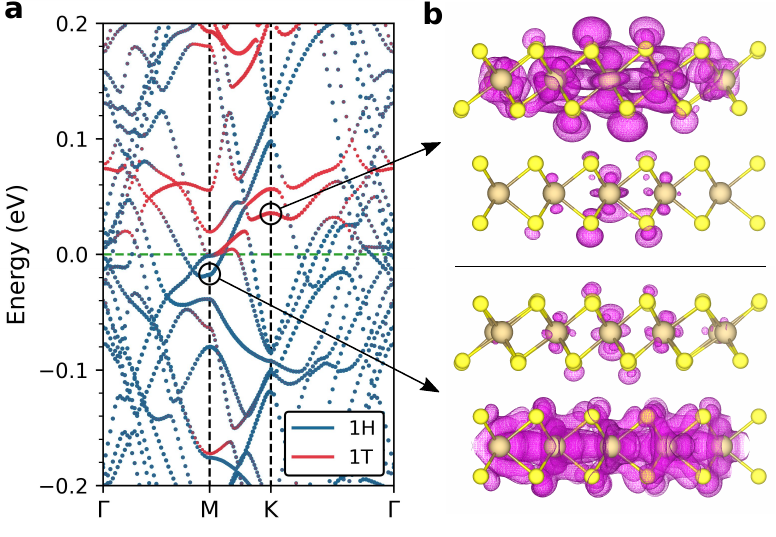}
	\caption{
	(a) Band structure of pristine 1T/1H bilayer. Contributions from each layer are given, with the 1T layer shown in red and the 1H layer shown in blue. The green dashed line shows the Fermi level.  (b) Charge densities of the selected points in the band structure. Brown spheres show tantalum, while the yellow spheres show sulfur. Isosurface values are set to 0.0002 $e$/\AA$^3$. 
	}
	\label{Fig11}
\end{figure}

Lastly, we wish to comment on an important aspect of the charge transfer analysis related to hybridization between states in the 1T and 1H layers. In Fig.~\ref{Fig11}(a), the band structure of the pristine 1T/1H bilayer is shown, with contributions from each layer indicated by different colors. In Fig.~\ref{Fig11}(b), we plot the real-space charge densities of selected Bloch states, where a noticeable degree of hybridization between the two layers can be observed. By integrating the charge density of the wave functions within the 1T and 1H layers, we find that approximately 10\% of the charge density of Bloch states predominantly associated with one layer resides in the opposite layer. 

This hybridization is non-negligible. For example, if a sulfur vacancy were to donate two electrons into bands predominantly associated with the 1T layer, about 0.2 of these electrons would reside in the 1H layer. This value is comparable to the charge transfer magnitudes reported in this work for several defects. In the charge transfer analysis, this portion of electrons would therefore be counted as occupying the 1H layer, even though in this thought experiment the electrons were introduced entirely into the 1T-derived bands. This illustrates an intrinsic aspect of charge partitioning in hybridized systems that should be kept in mind, particularly when making comparisons with angle-resolved photoemission spectroscopy, STM experiments, or model-based approaches such as doped Hubbard calculations.

\section{Conclusion and Outlook}

4$H_b$-TaS$_2$ is an intriguing system because it couples superconductivity in the 1H layer with the Star-of-David charge density wave in the 1T layer. Interlayer charge transfer has been proposed as a key ingredient for topological superconductivity, and STM experiments have shown that defects can be created or removed, thereby opening the door to local tuning the charge transfer and the exotic properties of 4$H_b$-TaS$_2$. The dominant defects in this material have been classified as type~2 defects, but their precise identification has remained challenging. In this work, we adopt a holistic approach, simulating over 90 defect configurations to identify the most plausible scenarios. Based on this analysis, we identified three candidates for type~2 defects: (i)~sulfur vacancies in the 1H layer, where oxygen substitutions may explain the absence of $V_{\mathrm{S}}^{\mathrm{1T}}$ in STM, although simulations do not capture the observed dimness at positive bias; (ii)~Ta-on-S antisites; and (iii)~Ta interstitials. Notably, for both Ta-based defects, the energetically preferred lateral positions are precisely those that reproduce the experimental STM contrast across both bias polarities. 

Several experimental strategies could help distinguish these scenarios. The oxygen concentration in the cleaving station or STM chamber could be varied to test its effect on the type~1/type~2 defect ratio. Higher-resolution STM could resolve atomistic details within the Star-of-David clusters to distinguish pristine S sites from $O_{\mathrm{S}}$ substitutions as has been done in Ref.~\cite{Barja2019}. Cross-sectional scanning transmission electron microscopy (STEM) could identify Ta atoms above S sites for Ta-on-S antisites or within the van der Waals gap for Ta interstitials [see Fig.~\ref{Fig9}]. Notably, because these defects occur at a ~10\% frequency relative to the 13 Ta atoms in a cluster, the resulting signal from these specific Ta columns would be significantly weaker than a standard Ta column, requiring high-sensitivity imaging to detect.

Regardless of the specific nature of the type~2 defects, our calculations show that formation energies vary strongly with growth conditions [S-poor versus S-rich]. Although a complete assessment of defect abundances would require consideration of finite-temperature thermodynamics and kinetic effects, the calculated formation energies suggest the possibility of tuning charge transfer through changes in the chemical potential. In particular, Ta-on-S antisites strongly influence the interlayer charge transfer, potentially providing a route to modulate the electronic properties of 4$H_b$-TaS$_2$. To support further work, all input and output files from this study have been made publicly available in an online database~\cite{data_repository}. This systematic first-principles dataset facilitates the understanding of defect-induced charge transfer in 4$H_b$-TaS$_2$ and serves as a comprehensive reference for future theoretical, experimental, and data-driven investigations.

\section*{ACKNOWLEDGMENTS}
This work was conducted at the Center for Nanophase Materials Sciences (CNMS), which is a U.S. Department of Energy, Office of Science User Facility. S.M. was supported by the U.S. National Science Foundation (NSF) through the EPSCoR Research Fellows program under Award No. 2531722. This research used resources of the National Energy Research Scientific Computing Center (NERSC), a Department of Energy User Facility using NERSC award BES-ERCAP0035988. We also used the Expanse supercomputer at the San Diego Supercomputer Center through allocation PHY230093 from the Advanced Cyberinfrastructure Coordination Ecosystem: Services \& Support (ACCESS) program, which is supported by National Science Foundation Grants No. 2138259, No. 2138286, No. 2138307, No. 2137603, and No. 2138296. The authors gratefully acknowledge the computational resources provided by the Hyperion high performance computing cluster at the University of South Carolina. We also acknowledge the technical assistance and resources provided by Research Computing at the University of South Carolina (RRID:SCR-027488). The crystal growth (H.D.Z) was supported by the U.S. Department of Energy (DOE) under Grant No. DE-SC0020254.

This manuscript has been authored by UT-Battelle, LLC under Contract No. DE-AC05-00OR22725 with the U.S. Department of Energy. The United States Government retains and the publisher, by accepting the article for publication, acknowledges that the United States Government retains a nonexclusive, paid-up, irrevocable, worldwide license to publish or reproduce the published form of this manuscript, or allow others to do so, for United States Government purposes. The Department of Energy will provide public access to these results of federally sponsored research in accordance with the DOE Public Access Plan (http://energy.gov/downloads/doe-public-access-plan). ChatGPT (GPT-5.6 Luna) was used for generative text editing and language refinement.

S.K., T.B., and S.M. conceptualized the project and wrote the manuscript. S.K. performed the calculations under the supervision of T.B. and S.M.. W.Y., W.K. and A.P.L. conducted the STM experiments. H.D.Z. carried out the crystal growth.

The authors declare no conflict of interest.

\section*{DATA AVAILABILITY}
Our study encompasses over 90 defect configurations and includes calculations of defect formation energies, charge transfer, work functions, density of states, and simulated STM images, made publicly available through an online database \cite{data_repository}.

\bibliography{TaS2theory}
\end{document}